\documentclass[referee, pdflatex, sn-vancouver,Numbered]{sn-jnl}

\usepackage{graphicx}%
\usepackage{multirow}%
\usepackage{amsmath,amssymb,amsfonts}%
\usepackage{amsthm}%
\usepackage[title]{appendix}%
\usepackage{xcolor}%
\usepackage{textcomp}%
\usepackage{manyfoot}%
\usepackage{booktabs}%
\usepackage{algorithm}%
\usepackage{algorithmicx}%
\usepackage{algpseudocode}%
\usepackage{listings}%
\usepackage{float}
\usepackage{lmodern}
\usepackage{mathrsfs}
\usepackage{hyperref}

\begin{document}

\title[Article Title]{A Bayesian Mixture Model Approach to Examining Neighborhood Social Determinants of Health Disparities in Endometrial Cancer Care in Massachusetts}

\author[1]{\fnm{Carmen B.} \sur{Rodr\'iguez}}\email{crodriguezcabrera@g.harvard.edu}

\author[2]{\fnm{Stephanie M.} \sur{Wu}}\email{stephanie.wu@ucl.ac.uk}

\author[3,4,5]{\fnm{Stephanie} \sur{Alimena}}\email{SALIMENA@partners.org}

\author[6]{\fnm{Alecia J}\sur{McGregor}}\email{amcgregor@hsph.harvard.edu}

\author*[1]{\fnm{Briana JK} \sur{Stephenson}}\email{bstephenson@hsph.harvard.edu}

\affil*[1]{\orgdiv{Department of Biostatistics}, \orgname{Harvard T.H. Chan School of Public Health}, \city{Boston}, \state{Massachusetts}, USA}

\affil[2]{\orgdiv{Public Health Data Science, Division of Psychiatry}, \orgname{University College London},  \city{London}, {United Kingdom}}

\affil[3]{\orgdiv{Division of Gynecologic Oncology, Department of Obstetrics and Gynecology and Reproductive Biology }, \orgname{Brigham and Women's Hospital},  \city{Boston}, \state{Massachusetts}, USA}

\affil[4]{\orgname{Dana-Farber Cancer Institute}, \city{Boston}, \state{Massachusetts},USA}

\affil[5]{\orgname{Harvard Medical School}, \city{Boston}, \state{Massachusetts},USA}

\affil[6]{\orgdiv{Department of Health Policy and Management},\orgname{Harvard T.H. Chan School of Public Health}, \city{Boston}, \state{Massachusetts},USA}

\newpage
 \abstract{Many studies have examined social determinants of health (SDoH) independently, overlooking their interconnected nature. Our study uses a multidimensional approach to construct a neighborhood-level measure that explores how multiple SDoH jointly impact care received for endometrial cancer (EC) patients in Massachusetts (MA). Using 2015-2019 American Community Survey data, we implemented a Bayesian multivariate Bernoulli mixture model to identify neighborhoods with similar SDoH features in MA. Five neighborhood SDoH (NSDoH) profiles were derived and characterized: (1) advantaged non-Hispanic White; (2) disadvantaged racially/ethnically diverse, more renter-occupied housing with limited English proficiency;(3) working class, lower educational attainment; (4) racially/ethnically diverse and greater economic security and educational attainment; and (5) racially/ethnically diverse, more renter-occupied housing with limited English proficiency. Based on residential information, we assigned these profiles to EC patients in the Massachusetts Cancer Registry. We used these profile assignments as the primary exposure in a Bayesian logistic regression to estimate the odds of receiving optimal EC care, adjusting for patient-level sociodemographic and clinical characteristics. NSDoH profiles were not significantly associated with receiving optimal EC care. However, compared to patients assigned to Profile 1, patients in all other profiles had lower odds of receiving optimal care. Our findings demonstrate how a flexible model-based clustering approach can account for the interconnected and multidimensional nature of NSDoH in a practical and interpretable way. Deriving and geospatially mapping NSDoH profiles may allow for identifying areas of need and inform targeted public health interventions tailored to each neighborhood’s specific social determinants to improve healthcare delivery.}


\maketitle
\newpage
\section{Background}\label{sec1}

Social determinants of health (SDoH) are defined as the social or demographic components of one's environment that shape health outcomes and can impact an individual's access to care, treatment decisions, and overall patient experiences \citep{cooper_social_2024, coughlin_social_2019, kolak_quantification_2020}. These determinants, including socioeconomic status, experiences of racism or discrimination, and neighborhood conditions, can introduce bias in the healthcare system, potentially leading to a lack of guideline-concordant treatment \citep{gomez_impact_2015}, such as for endometrial cancer (EC), one of the most commonly occurring female cancers. Despite the existence of evidence-based treatment guidelines, disparities across the EC care continuum persist, particularly among ethnically and socioeconomically disadvantaged groups \citep{paskett_eliminating_2020, doll_endometrial_2023, doll_endometrial_2018, huang_impact_2020, rodriguez_guideline-adherent_2021, eakin_alarming_2023, makker_endometrial_2021, rodriguez_racial-ethnic_2021, karia_racial_2023}. This disparate relationship serves as our motivation to better understand how SDoH patterns are considered and characterized \citep{makker_endometrial_2021,eakin_alarming_2023}. 
A comprehensive analysis of SDoH that transcends race and ethnicity would provide researchers and policymakers with a deeper understanding of how the social environment can influence a patient's ability to engage with the healthcare system \citep{rodriguez_guideline-adherent_2021,rodriguez_racial-ethnic_2021}.

Characterization of SDoH is often conducted using non-standardized composite scores of proxy variables that correspond to different domains, such as income, education, and employment \citep{rethorn_quantifying_2020, jonnalagadda_using_2020}. At the geographic level, SDoH has been operationalized using area-based measures derived from publicly available data sources such as the United States (US) Census and the American Community Survey (ACS). Two commonly used measures are the Yost Index and the Area Deprivation Index (ADI) \citep{yost_socioeconomic_2001,singh_area_2003,berg_adi-3_2021}. The Yost Index, typically used in national cancer surveillance, operates under the assumption that SDoH is one-dimensional. It inputs seven area-level variables (median household income, education, proportion of households below the 150\% poverty line, median house value, median gross rent, unemployment rate, and working-class occupation). Summarized at the census tract or block level, these variables are clustered via a Principal Component Analysis (PCA), based on how much the respective variable explains the variation in the data\citep{yost_socioeconomic_2001,lafantasie_empirical_2022,boscoe_comparison_2021}. Referred to as the primary component, the socioeconomic status (SES) index is calculated for each areal unit based on the variable loadings on that component. Areal units are then ranked based on their SES scores and divided into quintiles, with quintile 1 representing the lowest SES and quintile 5 the highest. As a one-dimensional component for SES, a loss of information results because this measure focuses on the one or two variables that may be most relevant to classifying low or high SES, as opposed to considering the combination of the different variables and how their contributions may differ from one residential area to another.

The ADI, originally developed in 2002 to study general mortality disparities and hospitalizations in the US, expands the set of census-derived SDoH variables to 17. Previously, it operated under the same one-dimensional assumption as Yost, but now incorporates a three-factor model to accommodate multi-dimensionality \citep{kind_making_2018, boscoe_comparison_2021, berg_adi-3_2021, singh_area_2003}. Highly correlated variables are first clustered together and then defined along the following three domains: financial strength, economic hardship and inequality, and educational attainment \citep{berg_adi-3_2021}. As in the Yost Index, areal units are indexed based on how high they load on the three-factor domains. Higher percentile ranks indicate greater socioeconomic disadvantage. A limitation of the ADI is that the derived domains are treated independently from one another without acknowledging the ways in which they intersect. Boscoe et al. compared the Yost and ADI indices and found that, despite being derived from the same data and heavily influenced by financial domains, they differ significantly in terms of transparency for reproducibility, distribution, and sensitivity to missing data \citep{boscoe_comparison_2021}. Methodologically, these approaches share similarities in their use of principal components or factor analysis to reduce the number of highly correlated SDoH variables, but the domains derived are treated independently from one another and fall short of acknowledging the ways in which they intersect.

Recently, Kolak et al. introduced a new approach that further expands the list of census-derived SDoH variables for PCA, including variables like limited English proficiency \citep{kolak_quantification_2020}. These variables were reduced to four domains: socioeconomic advantage index, limited mobility index, urban core opportunity index, and mixed immigrant cohesion and accessibility index. Additionally, they clustered census tracts by characterizing them into profiles based on similar scoring across the four SDoH components using K-means clustering. While this approach does allow for clustering based on shared responses to the different domains, the clustering is based on a single summary score. Overlooking how each component-specific score contributes to the summary score ignores the heterogeneity present in area deprivation. Two neighborhoods could potentially be clustered together but experience different patterns of resource allocation, which alter the level of impact these residents may experience in regards to their access and engagement with the healthcare system.  New methodological approaches are imperative to accommodate the interconnectedness and multifactorial structure of SDoH, inclusive of factors pertinent to vulnerable populations that are at greater risk of health disparities \citep{boscoe_comparison_2021,kolak_quantification_2020,rethorn_quantifying_2020, zhang_analyzing_2024}.

Model-based clustering, specifically finite mixture models (FMM), are able to identify underlying patterns in heterogeneous populations from a wide set of interrelated variables. Lekkas et al. recently highlighted the methodological advantages of FMMs in neighborhood health research through the common application of Latent Class Analysis (LCA) and Latent Profile Analysis (LPA). These models are able to characterize neighborhood profiles based on a set of multivariate categorical or continuous data and examine their association with various health indicators \citep{lekkas_finite_2019}. Although LCA or LPA are well-suited to handle the multi-faceted nature of neighborhoods, typically, the number of latent patterns/profiles is unknown, and multiple model fitting and testing is required to determine the optimal number of classes. 
However, under a Bayesian nonparametric framework, we are able to rely on dynamic and data-driven algorithms to identify an interpretable and reproducible set of neighborhood SDoH (NSDoH) profiles with the added flexibility of incorporating prior information into parameter estimation
\citep{wade_bayesian_nodate}. These models have found great utility in nutrition, environmental health, and population genomics \citep{wu_derivation_2023, stephenson_identifying_2024,stephenson_empirically_2020, maduekwe_identifying_2023,wang_bayesian_2017,gao_nonparametric_2024, medvedovic_bayesian_2004}.
Yet, none have been applied to area-level exposures of SDoH.

 In this study, we aim to generate and characterize new NSDoH profiles to better understand their impact on EC care in Massachusetts. Given binarized aggregate proportions of area-level SDoH variables from the ACS, we will implement a Bayesian multivariate Bernoulli mixture model (MBMM) to cluster neighborhoods that share similar SDoH characteristics. Neighborhoods are established based on census tract identifiers, as defined by the  American Community Survey from 2015 to 2019. Once identified, these profiles will be assigned to patients from the Massachusetts Cancer Registry, based on their residential data, and analyzed to determine if the care received was associated with their residential neighborhood. 

We organize the remaining sections of this paper as follows. In section \ref{sec2}, we describe our methodological approach, introducing the MBMM. In section \ref{sec3}, we apply the MBMM to our census-tract level data and examine its association with EC treatment. In section \ref{sec4}, we discuss the results and implications of this new approach to the field.

\section{Methods}\label{sec2}

\subsection{Data Source}\label{sec2_1}

We obtained census tract (neighborhood) level aggregate data from the 2015-2019 five-year American Community Survey (ACS)  estimates for the state of Massachusetts (MA). The data contained a total of 1478 census tracts, which are hereafter referred to as neighborhoods, that included approximately 6.8 million persons. Based on previous research, we selected fourteen (14) variables as indicators of NSDoH \citep{boscoe_comparison_2021, kolak_quantification_2020}.  
Given that the selected NSDoH variables were aggregate proportions and the Bernoulli distribution models binary data, we dichotomized each variable based on whether that feature fell above $(x_{ij}=1)$ or below $(x_{ij}=0)$ the state median value (Table \ref{tab1}). Dichotomization using the median was chosen due to the skewed distribution of several variables, which would have made other cutoffs (e.g., mean or quantile-based groupings) less interpretable or potentially biased by extreme values (Supplementary Figure \ref{figA1}). The median provides a robust, non-parametric threshold that ensures a balanced comparison between neighborhoods with relatively higher and lower exposures.  Neighborhoods above the state median threshold were classified as high exposure; those below were classified as low exposure.  Data was extracted from respective ACS tables via the \texttt{tidycensus} package implemented in the R software environment \citep{tidycensus_walker}. 

The 14 variables selected for analysis fall within four thematic domains: (1)  housing conditions and 
resources \citep{mehdipanah_neighborhood_2017, bowen_housing_2016, munford2020owning}; (2) economic security \citep{berg_adi-3_2021, lindberg_combining_2022,broer_review_2019,lut_health_2021}, (3) educational attainment, and (4) social and community context \citep{kolak_quantification_2020, berg_adi-3_2021}. Housing conditions and resources are described by the proportion of neighborhood households that were renter-occupied, lacked access to a vehicle, lacked complete plumbing, and experienced household crowding. Economic security is described by the proportion of households earning below the state median family income, participating in federal assistance (e.g., food stamps or SNAP - Supplemental Nutrition Assistance Program), as well as the unemployment rate, working-class status, and female head of households.  Educational attainment is described by the proportion of residents with no high school diploma. Social and community context is described by the dynamic interplay between multi-ethnic communities and the social structures that shape daily life 
\citep{sentell_low_2012,twersky2024impact}. We include these indicators intentionally as they can be considered proxies for racism, xenophobia, and bias, which would impact a resident's access to resources, opportunities, and social standing \citep{lett2022conceptualizing, krieger1999comparing, chant2004dangerous}. These indicators include the proportion of households with limited English (EN) proficiency and the proportion of residents identifying as Hispanic/Latino, non-Hispanic Black, and non-Hispanic Asian. To maintain interpretability across NSDoH variables, variables were recoded to reflect indicators of greater socioeconomic disadvantage. For example, the median family income variable was recoded such that a value of $(x_{ij}=1)$ represents low income (i.e., below the state median).

\subsection{Multivariate Bernoulli Mixture Model (MBMM)}\label{sec2_2}

Neighborhood social determinants of health (NSDoH) profiles were identified using a fully Bayesian estimation of a multivariate Bernoulli mixture model (MBMM) \citep{papastamoulis_bayesbinmix_2017}. This model clusters a population based on shared responses to a set of binary exposures. Let $i\in \{1,...,n\}$ index an individual census tract, where $n$ is the total number of neighborhoods in Massachusetts. Let $\mathbf{x_i} = \{x_{i,1},..,x_{i,p}\}$ denote a vector of observed binary indicators, where $x_{i,j}=1$ denotes a high exposure to social determinant $j\in \{1,..,p\}$ in census tract $i$.  Let $K$ denote the number of NSDoH profiles in the population. In practice, this number is typically not known \textit{a priori}.
To determine an appropriate number of profiles in the model, we overfit the model with $K_{max}$ clusters,  which over-exceeds the true number of clusters. This approach, coupled with a prior that treats the number of clusters $K < K_{\text{max}}$ as an unknown parameter, allows the observed data to drive the number of nonempty clusters estimated at the end of the Bayesian sampling algorithm \citep{van_havre_overfitting_2015, wade_bayesian_nodate, papastamoulis_overfitting_2018}.The observed likelihood for the set of binary data $\mathbf{X} = \{0,1\}^{n \times p}$ under the MBMM is given by

\begin{equation}\label{mbmm}
\mathcal{L}(\boldsymbol{\pi}, \boldsymbol{\theta}|\mathbf{X}) = \prod_{i=1}^n \biggl\{\sum_{k=1}^K \pi_k \prod_{j=1}^p \theta_{j|k}^{x_{i,j}} (1- \theta_{j|k})^{1-x_{i,j}}\biggl\},
\end{equation}

\noindent
where $\boldsymbol{\pi} = \{\pi_1,...,\pi_K\}$ is the probability vector for membership of a given neighborhood into one of the $K$ NSDoH profiles. The probability matrix $\boldsymbol{\theta} = \{\theta_{1|1},..., \theta_{j|k}\}^{p \times K}$ summarizes the set of individual SDoH variables, where $\theta_{j|k}$ is the probability of a high exposure to variable $j$ given the neighborhood's assignment to NSDoH profile $k$.  

For estimation of these parameters, we augment the data by introducing a latent allocation variable $\boldsymbol{z_{i}}$, such that $P(z_i = k) = \pi_k$. Therefore, we consider the complete data $\{x_i, z_i\}$ likelihood of the MBMM for computation:
\begin{equation}\label{mbmmcl}
\mathcal{L}^c(\boldsymbol{\pi}, \boldsymbol{\theta}|\mathbf{X}, \mathbf{Z})=\prod_{i=1}^n \prod_{k=1}^K  \biggl\{ \pi_{k} \prod_{j=1}^p \theta_{j|k}^{x_{i,j}} (1- \theta_{j|k})^{1-x_{i,j}}\biggl\}^{\mathbb{I}(z_i=k)}
\end{equation}

\subsubsection{Posterior Computation}\label{sec2_2_1}

Model parameters are estimated via a Metropolis-coupled Markov chain Monte Carlo ($MC^{3}$) algorithm described and implemented by Panagiotis Papastamoulis and Magnus Rattray as the R package \texttt{BayesbinMix} \citep{papastamoulis_bayesbinmix_2017}. An allocation sampler is used to determine the most probable number of NSDoH profiles. This is based on the maximum a posteriori number of nonempty clusters from the posterior distribution,$K_{map}$. Once $K_{map}$ is determined, we calculate the posterior mean of each model parameter. These posterior estimates are then used to describe each NSDoH profile $k$, based on the posterior distribution of $\boldsymbol{\theta_{\cdot|k}}=(\theta_{1|k},\ldots, \theta_{p|k})$, and assigning each MA census tract to the profile with the greatest posterior probability of membership. 

We assume no prior knowledge of parameter values and initialize them using noninformative priors. To ensure flexibility in capturing the complex, heterogeneous patterns of census tract-level data, we set a relatively large upper bound on the number of clusters, $K_{\text{max}} = 50$, which intentionally exceeds the number of clusters we expect to identify in this data setting, which avoids underfitting.

The following noninformative priors were imposed on the model parameters to allow the data to drive estimation:
$$
K| K_{max} \sim \text{Poisson} (\lambda = 1) \text{ truncated on the set} \{1,..., K_{max}\}
$$
$$
\boldsymbol{\pi} \mid K \sim \text{Dirichlet}(\gamma_1, \dots, \gamma_{K}), \quad \text{where } \gamma_k = 1 \; \forall \; k.
$$
$$
\theta_{j|k}|K \sim Beta (\alpha,\beta)\quad \text{where } \alpha = 1 = \beta \; \forall \; j,k
$$

The generated MCMC samples were postprocessed using the Equivalence Classes Representatives (ECR) algorithm to overcome label-switching identifiability issues inherent in Bayesian mixture models \citep{Papastamoulis01012014}. 
Further details on model fit testing and other cluster postprocessing procedures are provided in \textbf{Supplemental Materials Section \ref{secA}}.

\subsubsection{Regression Analysis}\label{sec2_2_2}
Patient-level data was obtained from the Massachusetts Cancer Registry (MCR) of the Massachusetts Department of Public Health (MDPH). A total of 2412 records were collected for women, aged 18 years or older, and diagnosed with endometrial cancer between 2015 and 2017. Cases were identified using the International Classification of Diseases for Oncology, third edition (ICD-O3) primary site (C54.1) and morphology codes for endometrial carcinoma \citep{who2013icdo}.  The main outcome variable was completion of recommended treatment according to the National Comprehensive Cancer Network (NCCN) guidelines as defined in the year of treatment for each patient. These guidelines are based on tumor, grade, and stage, incorporating a combination of surgery, chemotherapy, and radiation as necessary. Adherence to NCCN treatment guidelines was treated as a dichotomous variable, where we defined optimal care as the patient receiving therapy following NCCN guidelines. Patients were assigned to a respective MBMM-derived NSDoH profile based on their residential data collected at the time of diagnosis. Associations were measured by adjusted odds ratio via a Bayesian logistic regression model, which also adjusted for patient-level clinical and sociodemographic characteristics at time of diagnosis (year of diagnosis, age, insurance status, and initial point of care facility type). Supplementary analysis was conducted to examine the association between NSDoH profiles and the type of initial care facility (academic medical center versus other facilities). More details about the guidelines and regression results are provided in \textbf{Supplemental Materials Section \ref{secB}}.

All codes for the MBMM, data analysis, and data wrangling, including  ACS datasets, are available on GitHub at \url{https://github.com/cbrodriguez01/ecbayesbinmix}. All statistical analyses were conducted using the R software environment (version 4.3.1), and R packages (not exhaustive) 
\texttt{tidycensus}, \texttt{ggplot2}, \texttt{tidyverse}, \texttt{table1}, \texttt{BayesBinMix}, \texttt{coda}, \texttt{readxl}, \texttt{stringr}, \texttt{brms}, \texttt{shiny}, \texttt{leaflet}, \texttt{tigris}, and \texttt{sf}. The MBMM computations were run on the Harvard University Faculty of Arts and Sciences Research Computing (FASRC) Cannon cluster.

\section{Results}\label{sec3}

\subsection{NSDoH Profiles Results} \label{sec3_1}

Table \ref{tab1} shows the NSDoH variables selected from the 2015-2019 ACS survey and clustered to characterize neighborhoods in MA. Neighborhoods were mostly composed of owner-occupied households and high proportions of residents with at least a high school degree. On average, across all neighborhoods, about 13\% of households received federal assistance (i.e., SNAP) in the past 12 months, and 52\% of the population aged 16 or older had non-white collar occupations.

\begin{figure}[H]
    \centering
\includegraphics[width= 1.1\linewidth]{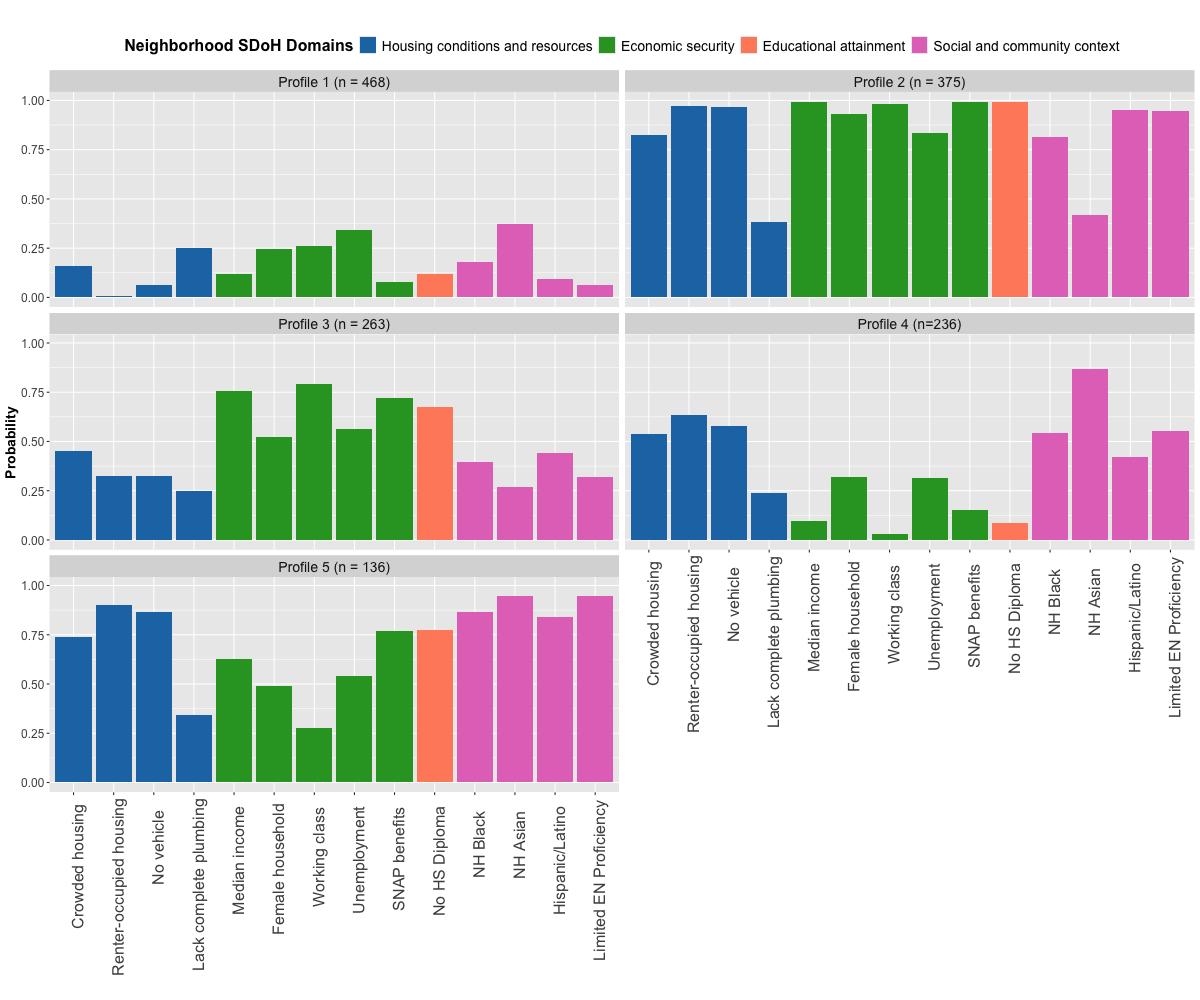}
    \caption{MBMM derived NSDoH profiles. Bars indicate the estimated posterior probability of high exposure to the NSDoH variable for a neighborhood, given assignment to an NSDoH profile. SDoH variables are ordered/colored by the NSDoH thematic domain. Abbreviations: high school(HS), EN(English), Non-Hispanic (NH), Supplemental Nutrition Assistance Program (SNAP). Shorthand names for NSDoH Profiles: 1) advantaged NHW; 2) disadvantaged racially/ethnically diverse (BHL+), more renter-occupied housing with limited EN proficiency;3)working class lower educational attainment; 4) racially/ethnically diverse (A+)  and greater economic security and educational attainment; 5)racially/ethnically diverse (ABHL+), more renter-occupied housing with limited EN proficiency.}
    \label{fig1}
\end{figure}

The model identified five NSDoH profiles, with a median profile assignment probability for neighborhoods of 0.95 (IQR: 0.18). Figure \ref{fig1} illustrates the NSDoH profile patterns derived from the MBMM. To facilitate interpretation, each profile was assigned a descriptive label reflecting the predominant NSDoH characteristics.  Profile 1, \emph{advantaged non-Hispanic White (NHW)}, represented about 32\% of MA neighborhoods with very low probabilities of exposure to poor NSDoH conditions across all domains of housing, employment, education, and social context (6.5\% - 37.4\%). Specifically, this profile could be described as containing neighborhoods mostly comprised of owner-occupied households, and median household incomes above or equal to the state's median. Profile 2, \emph{disadvantaged racially/ethnically diverse (BHL+; non-Hispanic Black (B) and Hispanic/Latino (HL)), more renter-occupied housing with limited EN proficiency}, represented the second largest cluster, containing 25\% MA neighborhoods. This profile exhibited characteristics of disadvantage and deprivation across all thematic domains. Neighborhoods in profile 2 had the highest probability of exposure to households with limited English proficiency (94.8\%), female head of households (93.1\%), household crowding (82.3\%), renter-occupied housing (97.3\%), no vehicle access (96.5\%), working class occupations (98.2\%), and higher proportions of all three ethnic minority groups, especially the proportion of non-Hispanic Black (81.4\%) and Hispanic/Latino residents (95.2\%). Profile 2 also contained neighborhoods with more residents exposed to low economic security (federal assistance participation - 99.3\%; median household income below the state's median - 99.3\%). Similar trends were found in the education domain, with all reporting no HS diploma (99.4\%), compared to all other NSDoH profiles.  Profile 3, \emph{working class lower educational attainment}, favored neighborhoods with high probability of no high school diploma (67.7\%), low probabilities of exposure to poor housing conditions and resources variables, and moderate proportions of ethnic minorities (26.8-43.9\%). Profile 4,\emph{racially/ethnically diverse (A+;non-Hispanic Asian (A)) and greater economic security and educational attainment}, shared similar low exposures to disadvantages in the economic security and educational attainment domains with Profile 1, but differed across housing conditions and resources and social and community context. For example, Profile 4 had more neighborhoods with a higher proportion of renter-occupied housing(63.6\%) and residents identifying as non-Hispanic Asians (NHA;86.7\%). Profile 5, \emph{racially/ethnically diverse (ABHL+), more renter-occupied housing with limited EN proficiency}, favored neighborhoods with high probabilities of exposure to crowded and renter-occupied housing, less than HS education, and ethnic minorities (84-95\%). Compared to Profile 2, Profile 5 had a higher proportion of non-Hispanic Asians and less exposure to poor economic security. 

\noindent

\subsection{Spatial Mapping of NSDoH} \label{sec3_2}
The MBMM profile assignments for each census tract were recorded and linked with MA census tract geographic data. We developed a web-based interactive map using \texttt{Shiny} to display the MBMM results, available at \url{https://mhn38j-carmen-rodriguez.shinyapps.io/nsdoh_profiles_app_urban/}. In this map, we also added an additional layer to provide further context to our MBMM defined profiles. Namely, the 2020 Urban Boundaries and Environmental Justice (EJ) Populations, obtained from the Massachusetts Bureau of Geographic Information Data Hub [\url{https://gis.data.mass.gov}]. The EJ data contains census block groups-- subdivisions within census tracts--across the state that meet one or more of the following criteria: (i) the annual median household income is not more than 65 \% of the statewide annual median household income; (ii) minorities comprise 40\% or more of the population; (iii) 25 percent or more of households lack English language proficiency.

Neighborhoods assigned to Profile 1 (advantaged NHW ) were found predominantly in surrounding areas outside of Boston (e.g., Berkshire, Hampshire, Essex, and Middlesex counties). Examining the overlaid urban boundaries reveals that most census tracts in this profile are located within urbanized areas. Additionally, the average proportion of non-Hispanic White individuals in these census tracts was 89\%, consistent with our profile results describing social and community context indicators. Neighborhoods assigned to Profile 2 (disadvantaged racially/ethnically diverse (BHL+), more renter-occupied housing with limited EN proficiency) were identified in areas with large immigrant populations (e.g., Lawrence, Methuen, Lynn, Chelsea, Dorchester, Revere, and Springfield) and a high average proportion of Hispanic/Latino residents. Additionally, most block groups that met the criteria for EJ populations, specifically regarding limited English proficiency in households and minority status, were located within the census tracts included in this profile. Neighborhoods in Profile 3(working class lower educational attainment) were primarily in areas west of Boston (e.g., Franklin and Worcester County). Neighborhoods in Profiles 4 (racially/ethnically diverse  (A+)  and greater economic security and educational attainment) and 5 (racially/ethnically diverse (ABHL+), more renter-occupied housing with limited EN proficiency) were primarily in areas with a known university presence and immediately surrounding Boston  (e.g., Cambridge, Allston, Brookline, Somerville). See  Supplementary Table \ref{tabA2} in the \textbf{Supplementary Materials Section \ref{secA}} for more information about the census tracts' population age and race-ethnicity distribution by NSDoH profile.

\subsection{NSDoH Profiles and Receipt of Optimal Care for EC}\label{sec3_3}

Supplementary Table \ref{tabB1} in the \textbf{Supplemental Materials Section \ref{secB}} shows sociodemographic and clinical information of the analytical sample of EC patients with endometroid histology who received optimal care or NCCN guidelines adherent treatment (82.3\%). These patients were predominantly non-Hispanic White (87.2\%), foreign-born (49.7\%), aged 50-64 (46.1\%), privately insured (46.8\%), diagnosed at stage I (91.5\%), and with grade 1 tumors (52.9\%). Patients were initially treated at large medical facilities (75.7\%), academic medical centers (38.7\%), and diagnosed by physicians specializing in family/internal medicine (47\%). 

Table \ref{tab2} details the distribution of patient demographics based on the NSDoH profile at which they resided at the time of EC diagnosis. Race-ethnicity, birthplace, insurance status at diagnosis, and initial point of care facility characteristics were significantly associated with NSDoH profiles. For example, among patients residing in neighborhoods belonging to NSDoH Profile 2, 49\%  were foreign-born, 42.6\% had Medicare or public insurance,  and 37.6\% were initially treated at academic medical centers. 

\begin{figure}[h]
    \centering
    \includegraphics[width=0.6\linewidth]{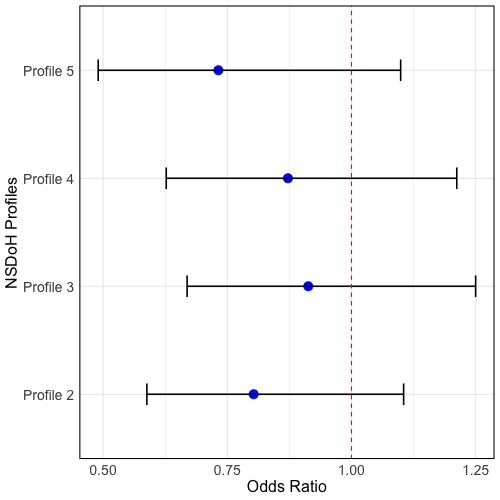}
    \caption{Adjusted odds of receiving optimal care by NSDoH profiles (Ref= NSDoH Profile 1; advantaged NHW) for patients diagnosed with EC from 2015-2019. Bars show 95\% Credible Intervals. Shorthand names for NSDoH Profiles: 1) advantaged NHW,
2) disadvantaged racially/ethnically diverse (BHL+), more renter-occupied housing with limited EN proficiency,
3) working class lower educational attainment, 
4) racially/ethnically diverse  (A+)  and greater economic security and educational attainment,
5) racially/ethnically diverse (ABHL+), more renter-occupied housing with limited EN proficiency.}
    \label{fig2}
\end{figure}

Figure \ref{fig2} summarizes the results of the regression analysis. Although not statistically significant, compared to patients who resided in neighborhoods belonging to the NSDoH Profile 1 (advantaged NHW), patients in the NSDoH Profile 2 (disadvantaged racially/ethnically diverse (BHL+), more renter-occupied housing with limited EN proficiency) had lower odds [OR = 0.80, 95\% Credible Interval (0.58,1.11)] of receiving optimal care after adjusting for year of diagnosis, age at diagnosis, insurance status at diagnosis, and initial type of care facility.

We also examined the relationship between NSDoH profiles and the type of initial care facility (academic medical center versus other facilities). The results, adjusted for year, age, and insurance status at diagnosis, are detailed in \textbf{Supplemental Materials Section \ref{secB}}. Compared to  NSDoH Profile 1 (advantaged NHW), patients who resided in NSDoH Profiles 4 (racially/ethnically diverse  (A+)  and greater economic security and educational attainment) and 5 (racially/ethnically diverse (ABHL+), more renter-occupied housing with limited EN proficiency
) were significantly associated with higher odds of receiving care at an academic medical center, while those in NSDoH Profile 3 (working class lower educational attainment) had lower odds, though not statistically significant.

\section{Discussion}\label{sec4}
We used data from the ACS to derive NSDoH profiles using a fully Bayesian multivariate Bernoulli mixture model. To our knowledge, this is the first study to apply a Bayesian mixture model to characterize neighborhood-level data in this context. This approach allowed us to estimate the appropriate number of profiles directly from the data, eliminating the need for post-hoc testing. Our model identified five NSDoH profiles, which we characterized using neighborhood SDoH variables belonging to four thematic domains: household conditions and resources, economic security, educational attainment, and social and community context. 

We used these profiles to conduct a regression analysis to examine the association of NSDoH profiles and optimal care for EC. Our findings, although not statistically significant, suggest some neighborhood profiles experience lower odds of optimal EC care, compared to patients in the advantaged NH White profile (Profile 1). Specifically, patients residing in neighborhood profiles with higher proportions of racially/ethnically diverse residents and greater housing and educational burden had lower odds of receiving optimal care. These findings are comparable to  Rodriguez et al. where they examined the association between neighborhood socioeconomic status (NSES) using the Yost Index and adherence to the NCCN guidelines from 2006-2015 \citep{rodriguez_racial-ethnic_2021}. Consistent with our findings, patients residing in the most disadvantaged neighborhoods with the lowest NSES had lower odds of receiving NCCN guideline-concordant care compared to those in the highest NSES group. 
In their study, 59.5\% of patients received treatment adhering to NCCN guidelines, with the lowest adherence seen among Black, Latina, and American Indian/Alaska Native women (57.1\%, 54.5\%, and 52.7\%, respectively). These proportions are lower than those in our study (82.3\%) but comparable to a study using the Women’s Health Initiative (WHI), where they found 80\% of patients received NCCN adherent treatment for EC \citep{felix2020guideline}. Our results focused specifically on Massachusetts, while their analysis covered the entire United States. Additionally,  similar to the Women's Health Initiative, our patient population was predominantly non-Hispanic White and limited to individuals engaged in the healthcare system. This demographic composition may further obscure disparities in access to and receipt of guideline-concordant care. Systemic barriers such as insurance status, provider bias, and structural racism have contributed to lower levels of healthcare engagement among some racial and ethnic minority populations, resulting in delayed screening, diagnosis, and more advanced disease at presentation \citep{najor2023disparities}. 
This is also important, as differences in standard treatment regimens may exist for some groups due to variations in the histologic subtypes and stages of endometrial cancer they present with \citep{Stewart_Nañez_Ayoola-Adeola_Chase_2023}.
Lastly, beyond the limitations of retrospective observational studies, the Massachusetts Cancer Registry lacks information on comorbidities, which may also impact differences in treatment adherence and recommendations. This could also explain the lack of guideline-concordant treatment rates for certain demographics.

Several studies have relied on the Yost index as a reliable measure of socioeconomic status at the neighborhood level \citep{ rodriguez_racial-ethnic_2021,rodriguez_guideline-adherent_2021, karia_racial_2023, goel_neighborhood_2023}. 
This measure of neighborhood disadvantage uses a subset of our variables, which yields a different narrative of SES that is not necessarily reflective of the socioeconomic barriers of healthcare access. The Yost index is derived via a one component PCA. This approach is a powerful and widely used method to reduce the dimensionality of high-dimensional datasets, but it suffers from interpretability issues because it outputs a linear combination of the original indicators \citep{jolliffe2016principal}. These results, by focusing on a composite score, could mask multidimensional deprivation patterns across neighborhoods. Our approach, which relied on a mixture model framework, provides a holistic way to identify patterns of multiple interrelated exposures jointly. This approach provides interpretable clusters of neighborhoods, making it easier to understand the heterogeneity of neighborhood data, as well as the defining characteristics of each profile. The Bayesian framework of the MBMM provides the flexibility of estimating the model parameters using prior information, including the parameter describing the number of profiles \citep{papastamoulis_bayesbinmix_2017, van_havre_overfitting_2015, miller_prior_2018}.  Utilizing a Bayesian framework allowed us to borrow information within and across other neighborhoods to improve the precision of our estimates, which improved the identifiability of our profiles.

This model is still met with limitations. First, model-based clustering is reflective of the data we use. In this study, we used 5-year estimates from the ACS from 2015-2019 for MA. Analysis using different ACS survey waves, different SDoH variables, or different geographies may yield different NSDoH profiles. For example, our results were based on binarized thresholds defined by median values for the state of Massachusetts. A different state would yield different cutoffs and, ultimately, different profiles. Second, the decision to dichotomize the SDoH variables was based on the skewed distributions found in the 2015-2019 cycles. This can sometimes result in a slight loss of information and reduced precision of the profile estimates since the full distribution of these variables was not considered. Future work should explore the development and implementation of a fully Bayesian multivariate beta mixture model that can flexibly accommodate bounded data, such as what we see in census-level data. Third, our analysis used aggregate data from the Census to construct NSDoH profiles. These variables are reflective of population-level characteristics rather than individual-level data. To avoid ecological fallacies, relationships observed at the aggregate level cannot be assumed to hold for individual EC patients residing within these neighborhoods. Lastly, consistent with other similar studies, our analysis is cross-sectional \citep{lekkas_finite_2019} and does not account for the changes in neighborhood demographics over time as a result of fair housing policies and gentrification. We focused our time frame on 2015-2019, which was based on the ACS 5-year estimates of data. Future studies should incorporate multiple survey cycles to assess how demographic shifts influence NSDoH profiles and, consequently, the access to and quality of healthcare and other outcomes for their residents. 

Bayesian mixture models offer promising applications for neighborhood-level data. Our approach enabled us to characterize and assign NSDoH profile patterns to neighborhoods in Massachusetts. Geospatial mapping of NSDoH profiles demonstrated how we can leverage these tools to identify areas for targeted interventions.  Using our NSDoH profiles to assess association with health outcomes, such as receipt of optimal care for EC, may give a more nuanced understanding of how SDoH overlap and co-occur within communities rather than in isolation to shape health experiences and outcomes. While this study focused on receipt of optimal care for EC patients, the derived profiles are translatable to a myriad of other outcomes and exploratory analyses. 

\section*{Competing interests}
The authors declare that they have no competing interests.
\section*{Acknowledgments}
We are grateful to everyone who contributed ideas or feedback, whether through brief conversations or informal discussions. We would also like to thank Dr. Giovanni Parmigiani for his valuable insight.

\section*{Funding}
This project was supported by the Harvard Data Science Initiative Bias$^2$ award given to the Stephenson Lab.  Dr. Stephenson is supported by NHLBI 1K01HL166442. Carmen B. Rodr\'iguez is supported by NIH T32CA009337. The funders had no role in the conceptualization, design, data analysis, decision to publish, or preparation of the manuscript.

\section*{Author contributions statement}
CBR and BJKS conceptualized and designed the study.
CBR performed all the data wrangling, analysis, interpretation of results, and drafting of the manuscript. SMW, SA, and AJM assisted in the interpretation of results. BJKS provided input and supervision over all data analysis and interpretation, as well as provided critical review and revisions to the manuscript. All authors read, contributed, and approved the final manuscript.

\section*{Data availability statement}
The R code used for this manuscript has been made publicly available in GitHub, \url{https://github.com/cbrodriguez01/ecbayesbinmix}. This also includes all the American Community Survey data we used for the multivariate Bernoulli mixture model, which is also publicly available at \url{https://www.census.gov/programs-surveys/acs/data.html}.
The data that support the findings from the regression analysis in this study are available from the Massachusetts Cancer Registry-Massachusetts  Department of Public Health (MCR-MDPH), but restrictions apply to the availability of MCR data, which were used under license for the current study, and so are not publicly available. Data are, however, available from the authors upon reasonable request and with permission of MCR-MDPH.

\begin{table}[h]
\footnotesize
\caption{Median estimate for Massachusetts for selected neighborhood social determinants of health and ACS 2015-2019 table identification.}
\label{tab1}
\begin{tabular}{lr}
    \toprule
    \textbf{NSDoH ACS Variable (ACS Table)} & \textbf{Median} \\
    \midrule
    \% of renter-occupied housing (DP04) & 33.85 \\
    \% Households without a motor vehicle (DP04) & 7.7 \\
    \% Crowding in household (DP04)$\dagger$ & 1.28 \\
    \% Occupied housing units without complete plumbing (DP04)$\ddagger$ & 0 \\
    Estimate median household income in the past 12 months (inflation-adjusted; B19013)$\S$ & 82265 \\
    \% Female Single-parent households with children younger than 18 (DP02) & 3.9 \\
    \%With Food Stamp/SNAP benefits in the past 12 months (DP03) & 7.7 \\
    \% Unemployed/unemployment rate (DP03) & 2.9 \\
    \% Employed population aged 16 years or older, working class (C24010) & 53.9 \\
    \% Population aged 25 years or older with no high school diploma  (DP02) & 6.5 \\
    \% Language other than English: Speak English less than ``very well''(DP02) & 5.8 \\
    \% Hispanics or Latinos (DP05) & 5.95 \\
    \% Non-Hispanic Black (DP05) & 2.8 \\
    \% Non-Hispanic Asian (DP05) & 3.2\\ 
    \bottomrule
\end{tabular}
\begin{tablenotes}
    \item[$\dagger$]{Occupied housing units with 1.01 to 1.50 and 1.51 or more occupants per room/ All occupied housing units for the same calendar year\citep{cdc_places}.}
    \item[$\ddagger$]{We binarized this variable differently because the median was 0. Therefore, census tracts for which there were no households lacking complete plumbing were categorized as ``0'',  and the remaining tracts were categorized as ``1'' to indicate lack of plumbing.}
    \item[$\S$]{Census tracts for which the median household income was $\ge$ to the state median were categorized as "0”, and remaining tracts were categorized as "1" to indicate lower income.}
\end{tablenotes}
\end{table}

\begin{table}[h]
\caption{Sociodemographic and Clinical Characteristics of Endometrial Carcinoma Cases Between 2015 – 2017 in the Massachusetts Cancer Registry (MCR) by NSDoH profiles (n=2412).} 
\label{tab2}
\centering
\begin{tabular}{llllll}
\hline
\textbf{} & \multicolumn{5}{c}{\textbf{Neighborhood SDoH Profiles}} \\
\multirow{2}{*}{\textbf{n(\%)}} & \multicolumn{1}{c}{\textbf{Profile 1}} & \multicolumn{1}{c}{\textbf{Profile 2}} & \multicolumn{1}{c}{\textbf{Profile 3}} & \multicolumn{1}{c}{\textbf{Profile 4}} & \multicolumn{1}{c}{\textbf{Profile 5}} \\
 & \multicolumn{1}{c}{\textbf{n= 952}} & \multicolumn{1}{c}{\textbf{n= 439}} & \multicolumn{1}{c}{\textbf{n=494}} & \multicolumn{1}{c}{\textbf{n=340}} & \multicolumn{1}{c}{\textbf{n=187}} \\ \hline
\textbf{Optimal care status} &  &  &  &  &  \\
Not optimal & 158 (16.6) & 88 (20.0) & 84 (17.0) & 59 (17.4) & 38 (20.3) \\
Optimal & 794 (83.4) & 351 (80.0) & 410 (83.0) & 281 (82.6) & 149 (79.7) \\
\textbf{Race-Ethnicity} &  &  &  &  &  \\
Non-Hispanic White & 910 (95.6) & 286 (65.1) & 459 (92.9) & 313 (92.1) & 135 (72.2) \\
Non-Hispanic Black & 7 ( 0.7) & 54 (12.3) & 12 ( 2.4) & 6 ( 1.8) & 20 (10.7) \\
Hispanic & 21 ( 2.2) & 13 ( 3.0) & 8 ( 1.6) & 13 ( 3.8) & 17 ( 9.1) \\
Other & 11 ( 1.2) & 82 (18.7) & 9 ( 1.8) & 7 ( 2.1) & 10 ( 5.3) \\
\textbf{Birthplace} &  &  &  &  &  \\
US-born & 439 (46.1) & 124 (28.2) & 166 (33.6) & 171 (50.3) & 68 (36.4) \\
Foreign-born & 458 (48.1) & 217 (49.4) & 298 (60.3) & 144 (42.4) & 84 (44.9) \\
Unknown &  55 (5.8) & 98 (22.3)& 30 ( 6.1)& 25 ( 7.4)& 35 (18.7) \\
\textbf{Year of diagnosis (y)} &  &  &  &  &  \\
2015 & 311 (32.7) & 160 (36.4) & 163 (33.0) & 120 (35.3) & 67 (35.8) \\
2016 & 345 (36.2) & 156 (35.5) & 172 (34.8) & 121 (35.6) & 61 (32.6) \\
2017 & 296 (31.1) & 123 (28.0) & 159 (32.2) & 99 (29.1) & 59 (31.6) \\
\textbf{Age at diagnosis (y)} &  &  &  &  &  \\
Younger than 50 & 69 ( 7.2) & 51 (11.6) & 39 ( 7.9) & 32 ( 9.4) & 23 (12.3) \\
50-64 & 433 (45.5) & 202 (46.0) & 236 (47.8) & 151 (44.4) & 89 (47.6) \\
65 or older & 450 (47.3) & 186 (42.4) & 219 (44.3) & 157 (46.2) & 75 (40.1) \\
\textbf{Insurance status at diagnosis} &  &  &  &  &  \\
Private & 463 (48.6) & 132 (30.1) & 204 (41.3) & 177 (52.1) & 88 (47.1) \\
Medicare & 403 (42.3) & 187 (42.6) & 203 (41.1) & 126 (37.1) & 67 (35.8) \\
Public/Government & 46 ( 4.8) & 78 (17.8) & 44 ( 8.9) & 19 ( 5.6) & 12 ( 6.4) \\
Other & 27 ( 2.8) & 38 ( 8.7) & 38 ( 7.7) & 12 ( 3.5) & 16 ( 8.6) \\
Not insured & 13 ( 1.4) & 4 ( 0.9) & 5 ( 1.0) & 6 ( 1.8) & 4 ( 2.1) \\
\textbf{Type of surgery received} &  &  &  &  &  \\
None & 28 ( 2.9) & 19 ( 4.3) & 20 ( 4.0) & 13 ( 3.8) & 11 ( 5.9) \\
Resection & 920 (96.6) & 415 (94.5) & 471 (95.3) & 327 (96.2) & 176 (94.1) \\
Other Surgery/Unknown & 4 ( 0.4) & 5 ( 1.1) & 3 ( 0.6) & 0 ( 0.0) & 0 ( 0.0) \\
\textbf{Type of radiation administered} &  &  &  &  &  \\
No radiation treatment & 691 (72.6) & 328 (74.7) & 367 (74.3) & 260 (76.5) & 143 (76.5) \\
External beam radiation therapy (EBRT) & 4 ( 0.4) & 4 ( 0.9) & 2 ( 0.4) & 3 ( 0.9) & 5 ( 2.7) \\
Brachytherapy & 174 (18.3) & 65 (14.8) & 80 (16.2) & 46 (13.5) & 18 ( 9.6) \\
Other & 83 ( 8.7) & 42 ( 9.6) & 45 ( 9.1) & 31 ( 9.1) & 21 (11.2) \\
\textbf{Chemotherapy status} &  &  &  &  &  \\
No & 833 (87.5) & 381 (86.8) & 443 (89.7) & 292 (85.9) & 152 (81.3) \\
Yes & 119 (12.5) & 58 (13.2) & 51 (10.3) & 48 (14.1) & 35 (18.7) \\
\textbf{Stage at diagnosis} &  &  &  &  &  \\
Stage I & 806 (84.7) & 361 (82.2) & 416 (84.2) & 278 (81.8) & 142 (75.9) \\
Stage II & 88 ( 9.2) & 42 ( 9.6) & 42 ( 8.5) & 39 (11.5) & 25 (13.4) \\
Stage III & 32 ( 3.4) & 11 ( 2.5) & 13 ( 2.6) & 11 ( 3.2) & 9 ( 4.8) \\
Stage IV & 26 ( 2.7) & 25 ( 5.7) & 23 ( 4.7) & 12 ( 3.5) & 11 ( 5.9) \\
\textbf{Grade at diagnosis} &  &  &  &  &  \\
Grade 1 & 442 (46.4) & 229 (52.2) & 253 (51.2) & 155 (45.6) & 87 (46.5) \\
Grade 2 & 340 (35.7) & 119 (27.1) & 158 (32.0) & 125 (36.8) & 61 (32.6) \\
Grade 3 & 170 (17.9) & 91 (20.7) & 83 (16.8) & 60 (17.6) & 39 (20.9) \\
\textbf{Initial point of care facility type} &  &  &  &  &  \\
Academic Medical Centers & 347 (36.4) & 165 (37.6) & 163 (33.0) & 168 (49.4) & 100 (53.5) \\
Community & 395 (41.5) & 125 (28.5) & 161 (32.6) & 96 (28.2) & 45 (24.1) \\
Specialty & 16 ( 1.7) & 6 ( 1.4) & 3 ( 0.6) & 3 ( 0.9) & 3 ( 1.6) \\
Teaching & 168 (17.6) & 126 (28.7) & 151 (30.6) & 72 (21.2) & 39 (20.9) \\
\textbf{Initial point of care facility size} &  &  &  &  &  \\
Small (\textless 100) & 35 ( 3.7) & 7 ( 1.6) & 15 ( 3.0) & 2 ( 0.6) & 2 ( 1.1) \\
Medium (100-299) & 257 (27.0) & 67 (15.3) & 97 (19.6) & 73 (21.5) & 32 (17.1) \\
Large (300+) & 660 (69.3) & 365 (83.1) & 382 (77.3) & 265 (77.9) & 153 (81.8) \\
\textbf{Initial point of care facility , doctor specialty} &  &  &  &  &  \\
Family/Internal medicine & 451 (47.4) & 198 (45.1) & 170 (34.4) & 193 (56.8) & 106 (56.7) \\
Hematology & 13 ( 1.4) & 2 ( 0.5) & 2 ( 0.4) & 5 ( 1.5) & 2 ( 1.1) \\
Gynecology and obstetrics & 201 (21.1) & 88 (20.0) & 115 (23.3) & 62 (18.2) & 16 ( 8.6) \\
Oncology & 136 (14.3) & 69 (15.7) & 84 (17.0) & 35 (10.3) & 24 (12.8) \\
Radiology & 28 ( 2.9) & 9 ( 2.1) & 16 ( 3.2) & 7 ( 2.1) & 4 ( 2.1) \\
Other specialty & 18 ( 1.9) & 5 ( 1.1) & 11 ( 2.2) & 10 ( 2.9) & 8 ( 4.3) \\
Unknown & 105 (11.0) & 68 (15.5) & 96 (19.4) & 28 ( 8.2) & 27 (14.4) \\ \hline
\end{tabular} 
\end{table}
 \clearpage
 
\setcounter{section}{0}
\section*{Web-based supplemental materials for ``A Bayesian Mixture Model Approach to Examining Neighborhood Social Determinants of Health Disparities in Endometrial Cancer Care in Massachusetts''}
\textbf{Carmen B Rodr\'iguez, Stephanie M Wu, Stephanie Alimena, Alecia J McGregor and Briana JK Stephenson}

\counterwithin{figure}{section}
\counterwithin{table}{section}

\section{Multivariate Bernoulli Mixture Model}\label{secA}

\begin{figure}[ht]
    \centering
    \includegraphics[width=1\linewidth]{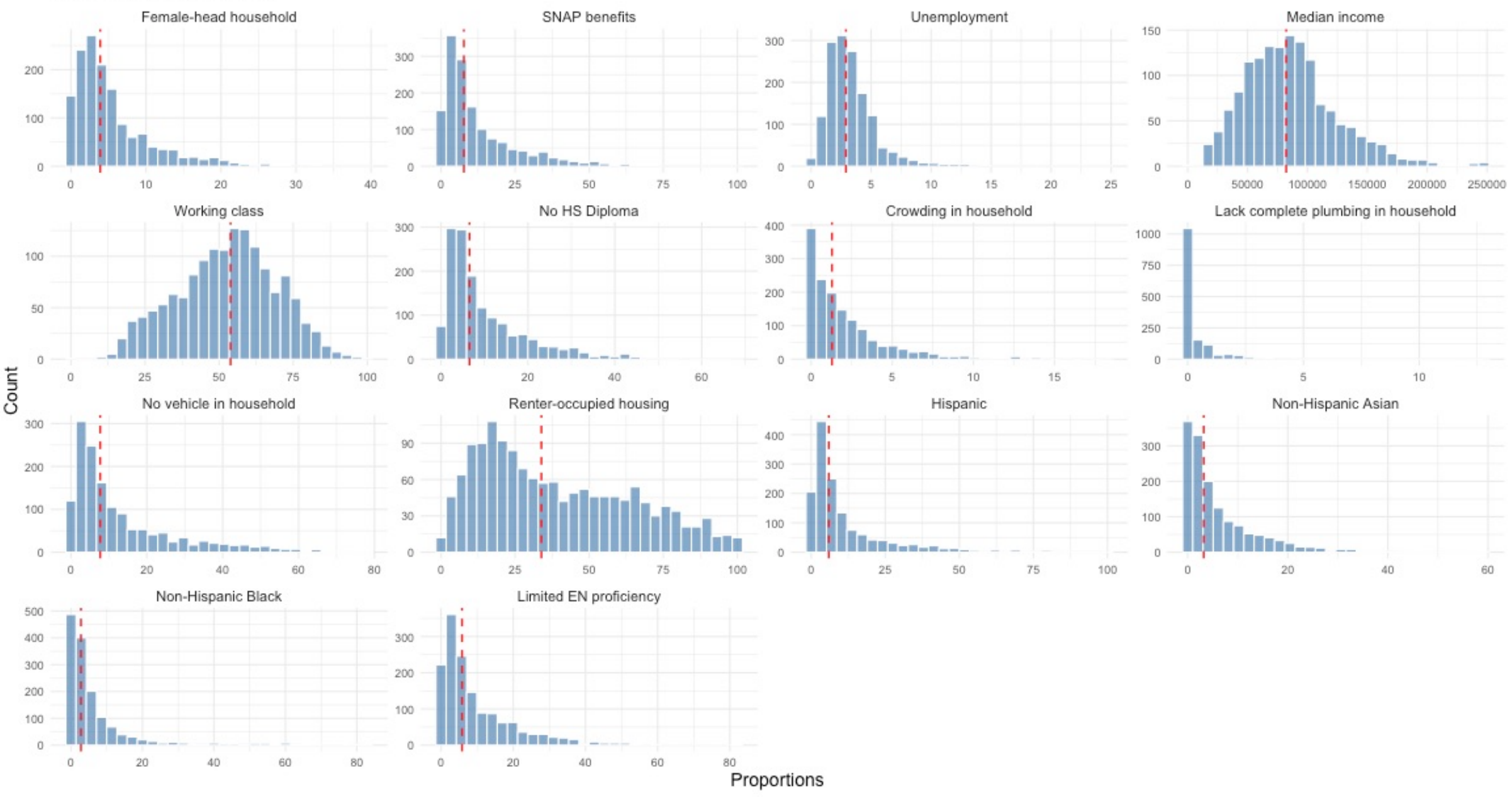}
    \caption{Distributions of NSDoH variables from ACS 2015-2019 5-year estimates for Massachusetts.}
    \label{figA1}
\end{figure}

Figure \ref{figA1} shows the distribution of all the NSDoH variables.  The distribution of these variables is skewed; therefore, as indicated by the vertical red line, we chose to dichotomize all variables for the MBMM model based on the median. We use the median because it is a robust non-parametric measure of central tendency unaffected by outliers. It ensures an equal data split, creating two comparison groups balanced in sample size. 

\begin{figure}[ht]
    \centering
    \includegraphics[width=0.70\linewidth]{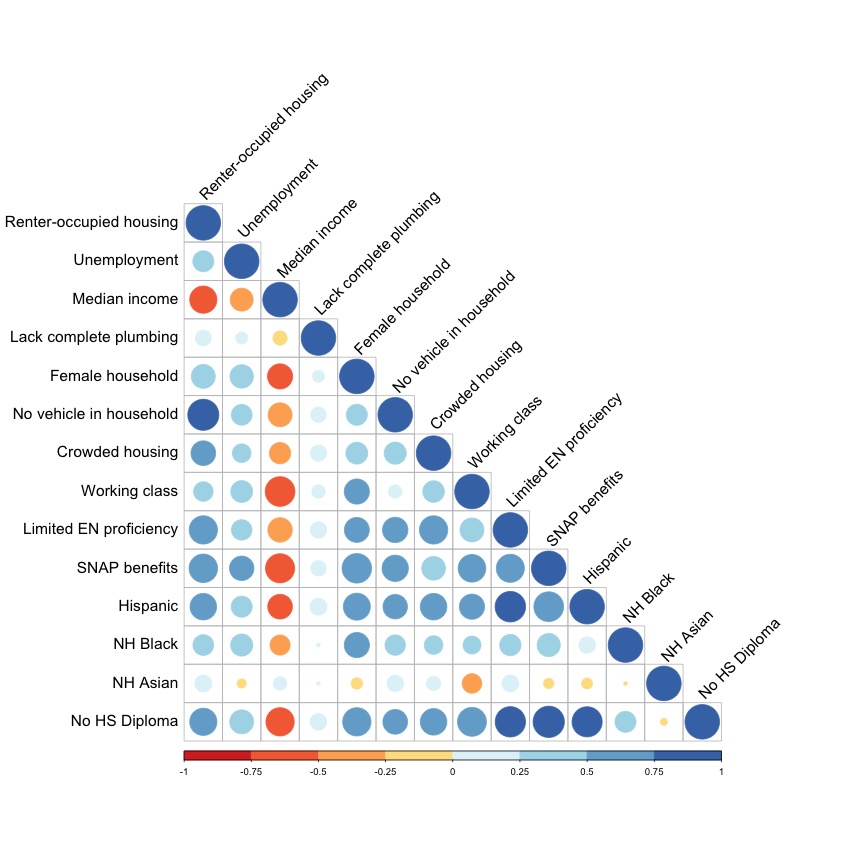}
    \caption{Pairwise Pearson's Correlations of NSDoH variables from ACS 2015-2019.}
    \label{figA2}
\end{figure}

Figure \ref{figA2} shows the pairwise correlations of the selected NSDoH variables. Given that we used a small geographic unit (i.e., census tracts), some variables are highly correlated. In the case of the MBMM, the variance-covariance matrix of a mixture component of independent Bernoulli distributions is not diagonal and does not assume independence between variables within clusters; therefore, this model can accommodate neighborhood-level data.

We assumed the observed data comes from a mixture of $K$ independent Bernoulli distributions.
Here, $K$ is the number of mixture components (i.e., clusters)\footnote{In the manuscript text, we refer to clusters as profiles.} included in the model, while the true number of clusters in the data is $K_0 < K$ and is determined by model estimation. That is, we assume that for a binary data matrix $\mathbf{X} = \{\mathbf{x_1},...\mathbf{x_n}\}$, each $\mathbf{x_i} = \{x_{i,1},...,x_{i,p}\} $ is such that $p(x_i)= \sum_{k=1}^K \pi_k \prod_{j=1}^p \theta_{j|k}^{x_{i,j}} (1- \theta_{j|k})^{1-x_{i,j}}$ as described in the methods Section 2.2, where $\boldsymbol{\pi} = \{\pi_1,...,\pi_K\}$ is the probability vector for the cluster assignments (i.e., the probability that a census tract belongs to cluster $k\in\{1,...,K\}$) and $\sum_{k=1}^K \pi_k = 1$, and the  probability matrix $\boldsymbol{\theta} = \{\theta_{j|k}\}^{p \times K}$ represents the probability of a high level of exposure to NSDoH variable $j$ for a census tract given assignment to cluster $k$.  Therefore, the observed likelihood for the specified model is 

\begin{equation}\label{mbmmsup}
\mathcal{L}(\boldsymbol{\pi}, \boldsymbol{\theta}|\mathbf{X}) = \prod_{i=1}^n \sum_{k=1}^K \pi_k \prod_{j=1}^p \theta_{j|k}^{x_{i,j}} (1- \theta_{j|k})^{1-x_{i,j}}
\end{equation}

We augment the data by introducing a latent allocation variable $\boldsymbol{z_{i}}$, such that $z_i = k$ means that census tract $i$ has been generated from the $k$-th cluster, and thus $P(z_i = k) = \pi_k$. For inference, we consider the complete data $\{x_i, z_i\}$ likelihood for the MBMM:
\begin{equation}\label{mbmmclsup}
\mathcal{L}^c(\boldsymbol{\pi}, \boldsymbol{\theta}|\mathbf{X}, \mathbf{Z})=\prod_{i=1}^n \prod_{k=1}^K  \biggl\{ \pi_{k} \prod_{j=1}^p \theta_{j|k}^{x_{i,j}} (1- \theta_{j|k})^{1-x_{i,j}}\biggl\}^{\mathbb{I}(z_i=k)}
\end{equation}

The Bayesian framework for computation relies on three key data components:  prior information about the model parameters, observed data likelihood, and posterior information. The prior information is defined as the first estimates of the distribution of the parameters before incorporating data through the Markov chain Monte Carlo (MCMC) sampler. As the sampler advances, both prior and observed data synergistically contribute to formulating insights about our target posterior distribution and updating information about the parameters using Bayes' Theorem. 

Estimation of the parameters for the MBMM  was performed using a Bayesian sampler described and implemented by Panagiotis Papastamoulis and Magnus Rattray as the R package \texttt{BayesbinMix} \citep{papastamoulis_bayesbinmix_2017}. The main function in the package is \texttt{coupleMetropolis()}, which embeds an allocation sampler \citep{nobile_bayesian_2007} with an unknown number of mixture components (i.e., a way to estimate the optimal number of components simultaneously) in a Metropolis-coupled Markov chain Monte Carlo ($MC^{3}$) algorithm. The $MC^{3}$ strategy is adopted to improve MCMC sampling by considering heated versions of the original target distribution. The function \texttt{coupleMetropolis()} takes as input the binarized observed data matrix $\mathbf{X}$, an upper bound on the number of clusters, defined in the function as $K_{max}$, prior information for all model parameters, and other inputs for computational efficiency. It then outputs the estimated posterior distribution of the model parameters $(\boldsymbol{\pi}, \boldsymbol{\theta})$ and the most probable number of NSDoH profiles defined as $(K_{map})$.For detailed information about this approach, please see their article \emph{BayesBinMix: an R Package for Model-Based Clustering of Multivariate Binary Data} \citep{papastamoulis_bayesbinmix_2017}.

We assume no prior knowledge of the initialization of the parameters. Consequently, all parameters are initialized with non-informative priors detailed below.  The true number of NSDoH profiles is unknown. Using the upper bound $K_{max}$( also referred to as $K$ in our model description above),  we impose an overfitted finite mixture model\citep{van_havre_overfitting_2015, papastamoulis_overfitting_2018}. We fit the model with a large upper bound on the number of clusters, $K = 50$, coupled with a prior on the number of clusters $K < K_{max}$, where the model treats $K$ as another unknown parameter, and allow a data-driven approach to estimating  $K$\citep{van_havre_overfitting_2015, wade_bayesian_nodate, papastamoulis_overfitting_2018}. We impose the following priors on model parameters:

$$
K| K_{max} \sim \text{Poisson} (\lambda = 1) \text{ truncated on the set} \{1,..., K_{max}\}
$$

For the other model parameters, we assume the following priors:

$$
\boldsymbol{\pi} \mid K \sim \text{Dirichlet}(\gamma_1, \dots, \gamma_{K}), \quad \text{where } \gamma_k = 1 \; \forall \; k.
$$

$$
\theta_{j|k}|K \sim Beta (\alpha,\beta)\quad \text{where } \alpha = 1 = \beta \; \forall \; j,k
$$

The following full conditional distributions were used to update the model parameters:
$$\boldsymbol{\pi}|K, \mathbf{Z} \sim \text{Dirichlet}\bigg(\gamma_1 + \sum_{i=1}^n \mathbb{I}(z_i=1),...,\gamma_K + \sum_{i=1}^n \mathbb{I}(z_i= K)\bigg)$$
$$\theta_{j|k}|K,\mathbf{X}, \mathbf{Z}  \sim \text{Beta}\bigg(\alpha + \sum_{i=1}^n \mathbb{I}(z_i=k)x_{i,j}, \beta + \sum_{i=1}^n \mathbb{I}(z_i=k) - \sum_{i=1}^n \mathbb{I}(z_i=k)x_{i,j}\bigg)$$ 
$$P(z_i = k|K, \mathbf{x}_i, \boldsymbol{\pi}, \boldsymbol{\theta}) \propto \pi_k \prod_{j=1}^p \theta_{j|k}^{x_{i,j}} (1- \theta_{j|k})^{1-x_{i,j}} $$

Given that they are smaller geographic units, some census tracts have missing information on some of the NSDoH variables (ranging from 14 to 24 census tracts with missing data across all variables). This model can handle missing data by imputing these values using the parameter's posterior mean estimates. The $MC^{3}$ for posterior computation was run for 15,000 iterations, with thinning every 10 iterations (i.e., retaining every 10th sample) and the first 5000 iterations removed as part of posterior samples post-processing. For the MCMC chain heating parameter denoted as $h_m, m=2,..., M$, where $M$ is the total number of parallel chains, we used incremental heating where the heat of the $m$'th chain is $h_m = 1/[1 + \Delta T \times (m-1)]$, and we tuned the parameter $\Delta T$ such that swaps between chains were accepted 20\%- 60\% of time \citep{altekar_parallel_2004}.  Through this process, we found that four (4) heated chains for the $MC^3$ algorithm produced good mixing, and for this type of data, smaller $\Delta T = 0.01$ works better. Posterior mean estimates were calculated from the remaining 1000 iterations collected from the sampler's output. The generated MCMC samples were postprocessed using the Equivalence Classes Representatives (ECR) algorithm to overcome label-switching identifiability issues inherent in Bayesian mixture models\citep{stephens_dealing_2000}. The most probable number of clusters ($K_{map}$) given the data was inferred, and the NSDoH profile assignment probabilities for each census tract were subsequently estimated after reordering with the ECR algorithm given $K_{map}$. Table \ref{tabA1} presents the median and interquartile range (IQR) of the assignment probabilities for census tracts to their most probable profile, while Figure \ref{figA2} illustrates the distribution of assignment probabilities across all NSDoH profiles given $K_{map}$. Some cluster assignment probabilities were relatively low; however, they still represented the highest probabilities among all cluster options for each census tract given $K_{map}$. These lower probabilities suggest the presence of unobserved heterogeneity, potentially driven by unmeasured NSDoH variables that we did not include in our model. Final clusters were qualitatively described based on thematic domains to define the NSDoH profiles, as discussed in Section 3.1 of the manuscript.

Additionally, we conducted sensitivity analyses using alternative priors (and combinations of priors) available in the software package. Specifically, we considered $K \sim Uniform(1, K_{max})$, and $\boldsymbol{\pi}|K \sim \text{Dirichlet}(\frac{1}{K_{max}},...,\frac{1}{K_{max}})$. Under these priors, we observed poorer MCMC mixing, as indicated by lower swap acceptance probabilities between heated chains. The estimated number of clusters consistently exceeded 10, and the cluster size distributions exhibited a combination of large and smaller clusters, including occasional singleton clusters. The emergence of singleton clusters is consistent with using a sparse Dirichlet prior on the mixing weights, which approximates a Dirichlet process and promotes the formation of smaller clusters.

 
\begin{table}[h]
\caption{MBMM estimated assignment probability of census tracts to the most probable profile.}
\label{tabA1}
\centering
\begin{tabular}{|l|c|}
\hline
\multicolumn{1}{|c|}{\textbf{NSDoH Profile}} & \textbf{Median (IQR)} \\ \hline
Profile 1 & 0.946 (0.161) \\ \hline
Profile 2 & 0.972 (0.0581) \\ \hline
Profile 3 & 0.923 (0.200) \\ \hline
Profile 4 & 0.874 (0.233) \\ \hline
Profile 5 & 0.865 (0.259) \\ \hline
\end{tabular}
\end{table}
\begin{figure}[!ht]
    \centering
    \includegraphics[width=0.88\linewidth]{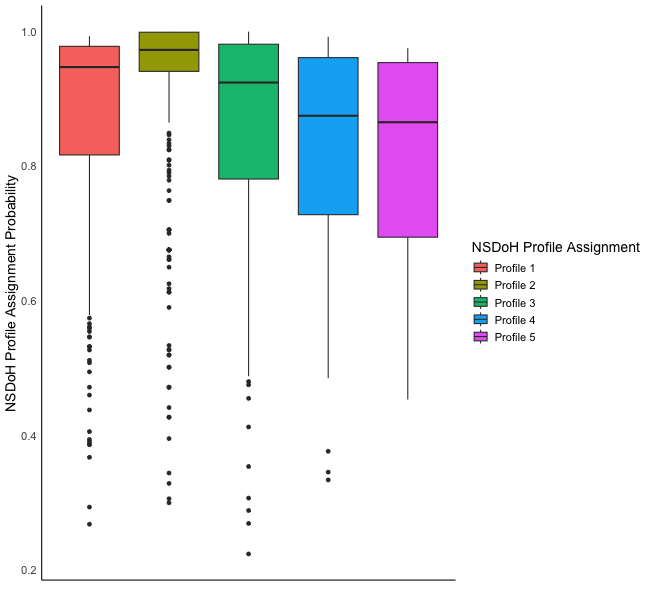}
    \caption{Distribution of assignment probabilities for each NSDoH profile.}
    \label{figA3}
\end{figure}

Table \ref{tabA2} shows the distribution of age and race/ethnicity within the census tracts used to further characterize NSDoH profiles in Section 3.2 of the manuscript.

\begin{table}[ht]
\caption{Distribution of population characteristics within census tracts in a given NSDoH Profile.}
\label{tabA2}
\begin{tabular}{|l|ccccc|}
\hline
& \multicolumn{5}{c|}{\textbf{Neighborhood SDoH Profiles}}  \\ \hline
& \multicolumn{1}{c|}{\textbf{Profile 1}} & \multicolumn{1}{c|}{\textbf{Profile 2}} & \multicolumn{1}{c|}{\textbf{Profile 3}} & \multicolumn{1}{c|}{\textbf{Profile 4}} & \textbf{Profile 5} \\ \hline
\textbf{Mean (SD)}      & \multicolumn{1}{c|}{\textbf{n = 468}}   & \multicolumn{1}{c|}{\textbf{n =  375}}  & \multicolumn{1}{c|}{\textbf{n= 263}}    & \multicolumn{1}{c|}{\textbf{n = 236}}   & \textbf{n = 136}   \\ \hline
Census tract population & \multicolumn{1}{c|}{5060 (1860)}        & \multicolumn{1}{c|}{4220 (1710)}        & \multicolumn{1}{c|}{4520 (1800)}        & \multicolumn{1}{c|}{4500 (1810)}        & 4770 (1660)        \\ \hline
Median Age              & \multicolumn{1}{c|}{45.6 (6.05)}        & \multicolumn{1}{c|}{35.3 (5.29)}        & \multicolumn{1}{c|}{43.3 (6.88)}        & \multicolumn{1}{c|}{38.0 (6.98)}        & 34.6 (5.79)        \\ \hline
Ages 20-24              & \multicolumn{1}{c|}{5.35 (4.94)}        & \multicolumn{1}{c|}{8.10 (5.15)}        & \multicolumn{1}{c|}{7.19 (6.98)}        & \multicolumn{1}{c|}{7.78 (5.95)}        & 10.8 (9.50)        \\ \hline
Ages 25-34              & \multicolumn{1}{c|}{9.12 (3.30)}        & \multicolumn{1}{c|}{16.3 (4.66)}        & \multicolumn{1}{c|}{12.3 (3.93)}        & \multicolumn{1}{c|}{20.4 (12.1)}        & 24.0 (9.04)        \\ \hline
Ages 35-44              & \multicolumn{1}{c|}{11.4 (2.88)}        & \multicolumn{1}{c|}{13.1 (3.34)}        & \multicolumn{1}{c|}{11.4 (3.18)}        & \multicolumn{1}{c|}{12.8 (3.17)}        & 12.8 (4.06)        \\ \hline
Age 45-54               & \multicolumn{1}{c|}{15.4 (3.15)}        & \multicolumn{1}{c|}{12.5 (3.14)}        & \multicolumn{1}{c|}{13.6 (3.42)}        & \multicolumn{1}{c|}{12.5 (4.22)}        & 10.7 (3.93)        \\ \hline
Age 55-59               & \multicolumn{1}{c|}{8.29 (2.02)}        & \multicolumn{1}{c|}{6.06 (1.93)}        & \multicolumn{1}{c|}{7.91 (2.69)}        & \multicolumn{1}{c|}{6.09 (2.44)}        & 5.48 (2.34)        \\ \hline
Ages 60-64              & \multicolumn{1}{c|}{7.66 (2.30)}        & \multicolumn{1}{c|}{5.44 (2.15)}        & \multicolumn{1}{c|}{7.38 (2.40)}        & \multicolumn{1}{c|}{5.47 (2.32)}        & 5.10 (2.20)        \\ \hline
Ages 65-74              & \multicolumn{1}{c|}{11.4 (4.11)}        & \multicolumn{1}{c|}{7.15 (2.68)}        & \multicolumn{1}{c|}{10.7 (3.92)}        & \multicolumn{1}{c|}{8.55 (4.93)}        & 7.21 (3.00)        \\ \hline
Ages 75-84              & \multicolumn{1}{c|}{5.51 (2.57)}        & \multicolumn{1}{c|}{3.59 (1.96)}        & \multicolumn{1}{c|}{5.42 (2.52)}        & \multicolumn{1}{c|}{4.35 (2.14)}        & 3.60 (1.88)        \\ \hline
Age 85 or older         & \multicolumn{1}{c|}{2.55 (1.81)}        & \multicolumn{1}{c|}{1.74 (1.30)}        & \multicolumn{1}{c|}{2.72 (1.95)}        & \multicolumn{1}{c|}{2.28 (1.93)}        & 2.02 (1.75)        \\ \hline
Non-Hispanic White      & \multicolumn{1}{c|}{89.0 (7.05)}        & \multicolumn{1}{c|}{41.9 (24.9)}        & \multicolumn{1}{c|}{82.0 (14.6)}        & \multicolumn{1}{c|}{74.7 (10.5)}        & 57.0 (13.2)        \\ \hline
Non-Hispanic Black      & \multicolumn{1}{c|}{1.69 (2.22)}        & \multicolumn{1}{c|}{16.2 (19.0)}        & \multicolumn{1}{c|}{4.58 (6.93)}        & \multicolumn{1}{c|}{4.30 (4.93)}        & 11.1 (9.91)        \\ \hline
Non-Hispanic Asian      & \multicolumn{1}{c|}{4.07 (4.99)}        & \multicolumn{1}{c|}{5.63 (9.05)}        & \multicolumn{1}{c|}{2.75 (4.12)}        & \multicolumn{1}{c|}{11.4 (8.04)}        & 14.3 (9.65)        \\ \hline
Hispanic or Latino      & \multicolumn{1}{c|}{3.14 (2.58)}        & \multicolumn{1}{c|}{31.9 (22.4)}        & \multicolumn{1}{c|}{7.61 (9.18)}        & \multicolumn{1}{c|}{6.28 (3.86)}        & 14.0 (9.48)        \\ \hline
\end{tabular}
\begin{tablenotes}
\item Shorthand names for NSDoH Profiles: 1) advantaged non-Hispanic White,
2) disadvantaged racially/ethnically diverse (BHL+; non-Hispanic Black (B) and Hispanic/Latino (HL)), more renter-occupied housing with limited EN proficiency,
3) working class lower educational attainment, 
4) racially/ethnically diverse  (A+; non-Hispanic Asian (A))  and greater economic security and educational attainment,
5) racially/ethnically diverse (ABHL+), more renter-occupied housing with limited EN proficiency.
\end{tablenotes}
\end{table}

\clearpage
\section{Regression Analysis Additional}\label{secB}

\begin{figure}[ht]
    \centering
    \includegraphics[width=0.88\linewidth]{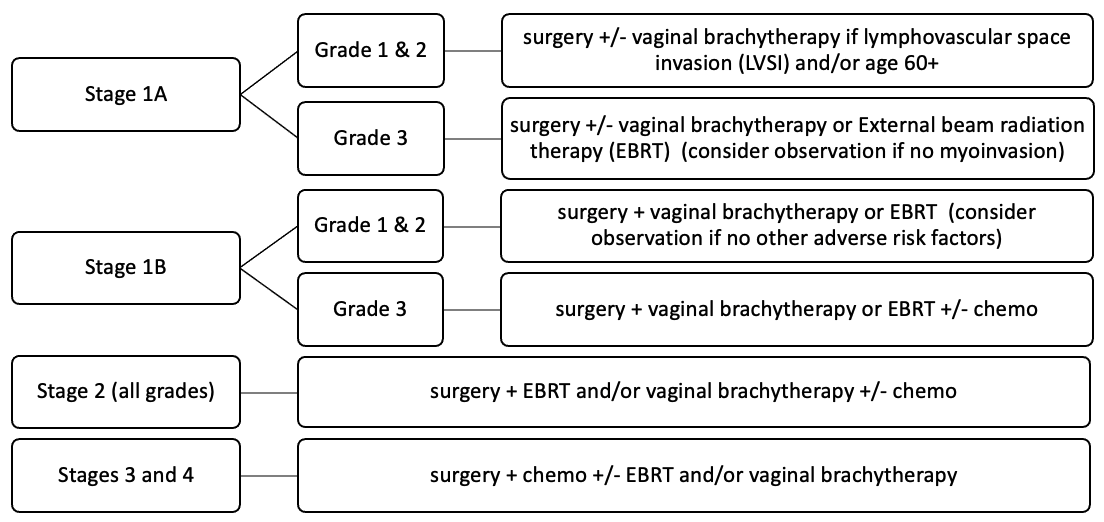}
    \caption{National Comprehensive Cancer Network guidelines example from 2020.}
    \label{figB1}
\end{figure}

The main outcome is optimal care, defined as adherence to National Comprehensive Cancer Network (NCCN) guidelines. These guidelines recommend a combination of therapies depending on the stage and grade of the tumor (Figure \ref{figB1}). We used year-specific guidelines spanning the study period due to gradual changes in NCCN treatment guidelines.  We determined optimal care by comparing the treatment received and the treatment recommended by NCCN. For example, we know that surgery is the first course of treatment, and thus if a woman received surgery alone or surgery and other additional therapies, as shown in Figure \ref{figB1}, based on the stage and grade of their tumor, then she is classified as having received optimal care.
We combined all the corresponding data to create a binary outcome variable, where: 
\begin{equation*}
Y_i = \begin{cases}
1 &\text{Received optimal care: patient received treatment following NCCN guidelines}\\
0 &\text{Did not receive optimal care}
\end{cases}
\end{equation*}

Patient's sociodemographic characteristics included age at diagnosis (younger than 50 years old, 50-64 years, or 65 and older), health insurance status at diagnosis (private, Medicare, public/government, uninsured/other), nativity ( U.S vs. foreign-born), and race/ethnicity (Hispanic, Non-Hispanic White, Non-Hispanic Black, Other race/ethnicity). Clinical characteristics included summary stage (distant, localized, regional), FIGO-stage (I,II,III,IV) using translations from the Summary Stage 2018 Coding Manual, tumor grade (1,2,3), and year of diagnosis. Treatment variables included type of surgery( none, resection, tumor destruction, other/unknown), type of radiation (brachytherapy, external beam (EBRT), other), and chemotherapy status (no, yes), and dates for each corresponding procedure. Additionally, we included information on the type (academic medical centers, community, specialty, and teaching) and size of the facility at the initial point of cancer management.

Table \ref{tabB1} shows the distribution of patient characteristics by optimal care status defined as NCCN treatment adherence. For all the Bayesian logistic regression analyses, we used the \texttt{brms} package with default non-informative priors designed to have minimal influence on the results. The model was run with 2,000 iterations per chain, including a 500-iteration burn-in. Two chains were used, and they showed good mixing for all parameters, indicating no evidence of non-convergence.

\begin{table}[!h]
\caption{Sociodemographic and Clinical Characteristics of Massachusetts Cancer Registry Endometrial Cancer Cases from 2015-2017 by Optimal Care Status (n=2412).}
\label{tabB1}
\begin{tabular}{|l|c|c|}
\hline
 & \textbf{\begin{tabular}[c]{@{}c@{}}Received \\ optimal care \\ (n=1985)\end{tabular}} & \textbf{\begin{tabular}[c]{@{}c@{}}Did not receive \\ optimal care\\ (n= 427)\end{tabular}} \\ \hline
\textbf{Race-Ethnicity} & \multicolumn{1}{l|}{} &  \\ \hline
Non-Hispanic White & 1739 (87.6) & 364 (85.2) \\ \hline
Non-Hispanic Black & 75 ( 3.8) & 24 ( 5.6) \\ \hline
Hispanic & 58 ( 2.9) & 14 ( 3.3) \\ \hline
Other & 99 ( 5.0) & 20 ( 4.7) \\ \hline
\textbf{Birthplace} & \multicolumn{1}{l|}{} &  \\ \hline
US-born & 796 (40.1) & 172 (40.3) \\ \hline
Foreign-born & 986 (49.7) & 215 (50.4) \\ \hline
Unknown & 203 (10.2) & 40 ( 9.4) \\ \hline
\textbf{Year of diagnosis (y)} & \multicolumn{1}{l|}{} &  \\ \hline
2015 & 667 (33.6) & 154 (36.1) \\ \hline
2016 & 723 (36.4) & 132 (30.9) \\ \hline
2017 & 595 (30.0) & 141 (33.0) \\ \hline
\textbf{Age at diagnosis (y)} & \multicolumn{1}{l|}{} &  \\ \hline
Younger than 50 & 181 ( 9.1) & 33 ( 7.7) \\ \hline
50-64 & 954 (48.1) & 157 (36.8) \\ \hline
65 or older & 850 (42.8) & 237 (55.5) \\ \hline
\textbf{Insurance Status at Diagnosis} & \multicolumn{1}{l|}{} &  \\ \hline
Private & 928 (46.8) & 136 (31.9) \\ \hline
Medicare & 769 (38.7) & 217 (50.8) \\ \hline
Public/Government & 152 ( 7.7) & 47 (11.0) \\ \hline
Other & 110 ( 5.5) & 21 ( 4.9) \\ \hline
Not insured & 26 ( 1.3) & 6 ( 1.4) \\ \hline
\textbf{Type of surgery received} & \multicolumn{1}{l|}{} &  \\ \hline
None & 0 ( 0.0) & 91 (21.3) \\ \hline
Resection & 1983 (99.9) & 326 (76.3) \\ \hline
Other/Unknown & 2 ( 0.1) & 10 ( 2.3) \\ \hline
\textbf{Type of radiation administered} & \multicolumn{1}{l|}{} &  \\ \hline
No radiation treatment & 1558 (78.5) & 231 (54.1) \\ \hline
External beam (EBRT) & 14 ( 0.7) & 4 ( 0.9) \\ \hline
Brachytherapy & 367 (18.5) & 16 ( 3.7) \\ \hline
Other & 46 ( 2.3) & 176 (41.2) \\ \hline
\textbf{Chemotherapy status} & \multicolumn{1}{l|}{} &  \\ \hline
No & 1828 (92.1) & 273 (63.9) \\ \hline
Yes & 157 ( 7.9) & 154 (36.1) \\ \hline
\textbf{Stage at diagnosis} & \multicolumn{1}{l|}{} &  \\ \hline
Stage I & 1816 (91.5) & 187 (43.8) \\ \hline
Stage II & 54 ( 2.7) & 182 (42.6) \\ \hline
Stage III & 62 ( 3.1) & 14 ( 3.3) \\ \hline
Stage IV & 53 ( 2.7) & 44 (10.3) \\ \hline
\textbf{Grade at diagnosis} & \multicolumn{1}{l|}{} &  \\ \hline
Grade 1 & 1050 (52.9) & 116 (27.2) \\ \hline
Grade 2 & 670 (33.8) & 133 (31.1) \\ \hline
Grade 3 & 265 (13.4) & 178 (41.7) \\ \hline
\textbf{Initial point of care facility type} & \multicolumn{1}{l|}{} &  \\ \hline
Academic Medical Centers & 768 (38.7) & 175 (41.0) \\ \hline
Community & 693 (34.9) & 129 (30.2) \\ \hline
Specialty & 15 ( 0.8) & 16 ( 3.7) \\ \hline
Teaching & 469 (23.6) & 87 (20.4) \\ \hline
\textbf{Initial point of care facility size} & \multicolumn{1}{l|}{} &  \\ \hline
Small (< 100) & 39 ( 2.0) & 22 ( 5.2) \\ \hline
Medium (100-299) & 450 (22.7) & 76 (17.8) \\ \hline
Large (300+) & 1496 (75.4) & 329 (77.0) \\ \hline
\textbf{Initial point of care facility , doctor specialty} & \multicolumn{1}{l|}{} &  \\ \hline
Family/Internal medicine & 933 (47.0) & 185 (43.3) \\ \hline
Hematology & 12 ( 0.6) & 12 ( 2.8) \\ \hline
Gynecology \& obstetrics & 407 (20.5) & 75 (17.6) \\ \hline
Oncology & 305 (15.4) & 43 (10.1) \\ \hline
Radiology & 45 ( 2.3) & 19 ( 4.4) \\ \hline
Other specialty & 44 ( 2.2) & 8 ( 1.9) \\ \hline
Unknown & 239 (12.0) & 85 (19.9) \\ \hline
\end{tabular}
\end{table}

We also explored the association between NSDoH profiles and the type of facility where patients were diagnosed or received treatment. Due to data limitations, we could not distinguish between these scenarios. For this sub-analysis, we dichotomized the facility type as an academic medical center versus other types. Similarly to the optimal care analysis, we conducted a Bayesian logistic regression adjusted for age, year, and insurance status at diagnosis, with results in Table \ref{tabB2}. Even after adjustment, NSDoH profiles 4 (racially/ethnically diverse  (A+)  and greater economic security and educational attainment) and 5 (racially/ethnically diverse (ABHL+), more renter-occupied housing with limited EN proficiency)  were linked to higher odds of receiving care at academic centers. The interactive map in Section 3.2 shows that neighborhoods in profiles 4 and 5 are often located near universities, which are in proximity to academic medical centers.
In contrast, although not statistically significant,  patients residing in neighborhoods in NSDoH profile 3 (working class lower educational attainment) had lower odds of receiving care at an academic facility.  These patients were more likely to receive care at either a community or teaching hospital (Table \ref{tab2}), and from observing the interactive map in Section 3.2 for NSDoH profile 3, most neighborhoods may be located farther from academic medical centers, making it less convenient for residents to access these facilities. Despite lower probabilities of "no vehicle in the household"  in this profile, transportation challenges may still be a barrier, particularly for working-class populations (for which there is a high probability of 79\%) with limited flexibility in work schedules.

\begin{table}[!ht]
\footnotesize
\caption{Univariate and Multivariate Logistic Regression of Receiving Care at an Academic Medical Center Among Endometrial Cancer Cases Between 2015 and 2017 in the Massachusetts Cancer Registry (n=2412).}
\label{tabB2}
\begin{tabular}{|r|cc|cc|}
\hline
\multicolumn{1}{|c|}{\textbf{}} & \multicolumn{2}{c|}{\textbf{Model 1}} & \multicolumn{2}{c|}{\textbf{Model 2}} \\ \hline
\multicolumn{1}{|c|}{} & \multicolumn{1}{c|}{\textbf{OR}} & \textbf{95\% Credible Interval} & \multicolumn{1}{c|}{\textbf{OR}} & \textbf{95\% Credible Interval} \\ \hline
\textbf{Neighborhood SDoH Profile} & \multicolumn{4}{c|}{} \\ \hline
Profile 1  & \multicolumn{2}{c|}{Referent} & \multicolumn{2}{c|}{Referent} \\ \hline
Profile 2 & \multicolumn{1}{c|}{1.049} & (0.835, 1.324) & \multicolumn{1}{c|}{1.116} & (0.875, 1.424) \\ \hline
Profile 3 & \multicolumn{1}{c|}{0.860} & (0.674, 1.083) & \multicolumn{1}{c|}{0.883} & (0.702, 1.101) \\ \hline
Profile 4 & \multicolumn{1}{c|}{1.701} & (1.328, 2.181) & \multicolumn{1}{c|}{1.697} & (1.313, 2.203) \\ \hline
Profile 5 & \multicolumn{1}{c|}{2.011} & (1.482, 2.730) & \multicolumn{1}{c|}{2.033} & (1.464, 2.825) \\ \hline
\end{tabular}
\begin{tablenotes}
\item Model 1: Unadjusted model.
\item Model 2: Adjusted for age, year, and insurance status at diagnosis.
\item Shorthand names for NSDoH Profiles: 1) advantaged non-Hispanic White,
2) disadvantaged racially/ethnically diverse (BHL+), more renter-occupied housing with limited EN proficiency,
3) working class, lower educational attainment, 
4) racially/ethnically diverse  (A+)  and greater economic security and educational attainment,
5) racially/ethnically diverse (ABHL+), more renter-occupied housing with limited EN proficiency.
\end{tablenotes}
\end{table}

\clearpage
\bibliography{ECMBMM_finalrefs}

\end{document}